\documentclass[onecolumn, preprintnumbers, superscriptaddress, aps, prl, 12pt, floatfix, tightenlines]{revtex4-1}               

\usepackage{CJKutf8}
\usepackage{latexsym}
\usepackage{amstext}
\usepackage{amsfonts}
\usepackage{dsfont}
\usepackage{color}
\usepackage{amssymb}
\usepackage{amsmath}
\usepackage{relsize}
\usepackage{amsxtra}
\usepackage{verbatim}
\usepackage{hyperref}
\usepackage{graphicx}

\AtBeginDvi{\input{zhwinfonts}}
\begin{document}
\begin{CJK*}{UTF8}{zhkai}

\title{Possible magnetic-polaron-switched positive and negative \\
magnetoresistance in the GdSi single crystal}

\author{Hai-Feng Li} 
\email{h.li@fz-juelich.de; hfli2101@gmail.com} \affiliation{J$\ddot{u}$lich Centre for Neutron Science JCNS, Forschungszentrum J$\ddot{u}$lich GmbH, Outstation at Institut Laue-Langevin, Bo$\hat{\imath}$te Postale 156, F-38042 Grenoble Cedex 9, France} \affiliation{Institut f$\ddot{u}$r Kristallographie der RWTH Aachen University, D-52056 Aachen, Germany}
\author{Yinguo Xiao}
\affiliation{J$\ddot{u}$lich Centre for Neutron Science JCNS and Peter Gr$\ddot{u}$nberg Institut PGI, JARA-FIT, Forschungszentrum J$\ddot{u}$lich GmbH,
D-52425 J$\ddot{u}$lich, Germany}
\author{Berthold Schmitz}
\affiliation{J$\ddot{u}$lich Centre for Neutron Science JCNS and Peter Gr$\ddot{u}$nberg Institut PGI, JARA-FIT, Forschungszentrum J$\ddot{u}$lich GmbH,
D-52425 J$\ddot{u}$lich, Germany}
\author{J$\ddot{\texttt{o}}$rg Persson}
\affiliation{J$\ddot{u}$lich Centre for Neutron Science JCNS and Peter Gr$\ddot{u}$nberg Institut PGI, JARA-FIT, Forschungszentrum J$\ddot{u}$lich GmbH,
D-52425 J$\ddot{u}$lich, Germany}
\author{Wolfgang Schmidt}
\affiliation{J$\ddot{u}$lich Centre for Neutron Science JCNS, Forschungszentrum J$\ddot{u}$lich GmbH, Outstation at Institut Laue-Langevin,
Bo$\hat{\imath}$te Postale 156, F-38042 Grenoble Cedex 9, France}
\author{Paul Meuffels}
\affiliation{Peter Gr$\ddot{u}$nberg Institut PGI and JARA-FIT, Forschungszentrum J$\ddot{u}$lich GmbH, D-52425 J$\ddot{u}$lich, Germany}
\author{Georg Roth}
\affiliation{Institut f$\ddot{u}$r Kristallographie der RWTH Aachen University, D-52056 Aachen, Germany}
\author{Thomas Br$\ddot{\texttt{u}}$ckel}
\affiliation{J$\ddot{u}$lich Centre for Neutron Science JCNS and Peter Gr$\ddot{u}$nberg Institut PGI, JARA-FIT, Forschungszentrum J$\ddot{u}$lich GmbH,
D-52425 J$\ddot{u}$lich, Germany}

\begin{abstract}
\textbf{Magnetoresistance (MR) has attracted tremendous attention for possible technological applications. Understanding the role of magnetism in manipulating MR may in turn steer the searching for new applicable MR materials. Here we show that antiferromagnetic (AFM) GdSi metal displays an anisotropic positive MR value (PMRV), up to $\sim$ 415\%, accompanied by a large negative thermal volume expansion (NTVE). Around $T_\texttt{N}$ the PMRV translates to negative, down to $\sim$ -10.5\%. Their theory-breaking magnetic-field dependencies [PMRV: dominantly linear; negative MR value (NMRV): quadratic] and the unusual NTVE indicate that PMRV is induced by the formation of magnetic polarons in 5\emph{d} bands, whereas NMRV is possibly due to abated electron-spin scattering resulting from magnetic-field-aligned local 4\emph{f} spins. Our results may open up a new avenue of searching for giant MR materials by suppressing the AFM transition temperature, opposite the case in manganites, and provide a promising approach to novel magnetic and electric devices.}
\end{abstract}

\maketitle
\end{CJK*}


Magnetoresistance (MR), a change in electrical resistivity when an external magnetic field ($\mu_0H$) applied, occurs in metals, inorganic and organic semiconductors, and particularly close to an intermediate regime of the transformations of charge (insulator, metal) order and spin [paramagnetic (PM), ferromagnetic (FM)] order in thin manganites' films as a colossal negative MR value (NMRV) \cite{Santen1950, Jin1994, Shimakawa1996, Majumdar1998, Xiong2004}. The colossal MR (CMR) in manganites is accompanied by a shift of the transition temperature to a higher value by applied magnetic field so that a sharp peak appears in the MR values near the transition, e.g., the NMRV reaches $\sim$ -95\% at $\mu_0H$ = 15 T near $T_\texttt{c}$ = 240 K in La$_{0.85}$Sr$_{0.15}$MnO$_3$ \cite{Urushibara1995}. Although the double exchange interaction can qualitatively explain the CMR effect based only on the charge and spin degrees of freedom, the exact picture still remains elusive.

We investigate here the intermetallic GdSi that crystallizes in the FeB-type structure (\emph{Pnma}) and orders antiferromagnetically below $\sim$ 55 K \cite{Gladyshevskii1964, nagaki1990, Saito1996, Tung2005, Roger2006}. The vanished orbital momentum (\emph{L} = 0) naturally removes the crystal electric field (CEF) as well as its perturbation on magnetic interactions, which renders GdSi ideal for studying pure spin-magnetism-tuned transport. Meanwhile, there keep solely two main sources for the magnetovolume (MV) effect. One contribution is from the volume dependence of the Ruderman-Kittel-Kasuya-Yosida (RKKY) exchanges, which normally produces a negative MV (NMV) effect. The other is due to the spin-polarized itinerant moments in the conduction bands, resulting in a positive MV (PMV) effect. Therefore, the volume variation (NMV or PMV) with temperature in a magnetic state could be useful in determining the nature of its magnetic origin. Even after nearly 50 years of research on GdSi \cite{nagaki1990, Saito1996, Tung2005, Roger2006}, to our knowledge, there is little data on its temperature-dependent structural modifications and MR property. Here we report on detailed temperature-dependent powder diffraction, angular-dependent magnetic characteristic and transport studies of GdSi single crystals. We find anomalous anisotropic giant MR and spontaneous magnetostriction (MS) effects, in particular, an antiferromagnetic (AFM)-driven negative thermal volume expansion (NTVE) and a nontrivial positive to negative MR transition, which can be well understood by combining the magnetic tunnel of conduction electrons and the concept of magnetic polarons \cite{Heikes1964}, i.e., local short-range FM spin regimes \cite{Heikes1964}.

\begin{figure}
\centering \includegraphics[width = 0.48 \textwidth] {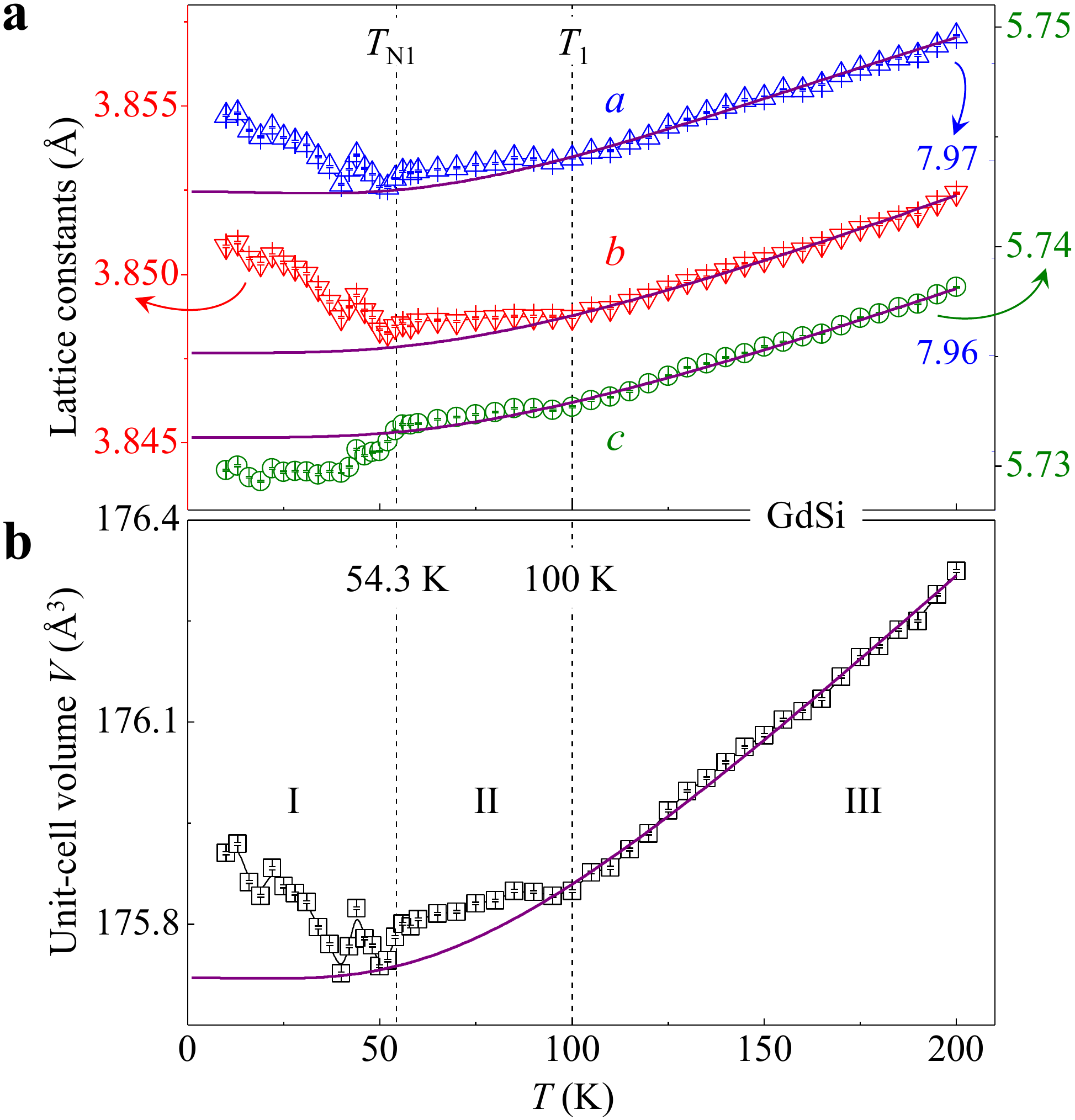}
\caption{\textbf{Temperature-dependent structural parameters.}
(a) Anisotropic character of the lattice-constants, \emph{a}, \emph{b} and \emph{c}, variation. (b) Anomalous unit-cell volume, \emph{V},
expansion with temperature in the \emph{Pnma} symmetry. The solid lines are theoretical estimates of the temperature-dependent structural parameters using
the Gr$\ddot{\texttt{u}}$neisen model with Debye temperature of $\theta_\texttt{D}$ = 340 K, which is the same as reported in Ref. \cite{Tung2005}. Upon warming, two appreciable anomalies display in the structural parameters at respective temperatures of $T_\texttt{N1}$ $\approx$ 54.3 K (0.06 T) and $T_1$ $\approx$ 100 K. Error bars in (a) and (b) are the standard deviation obtained from the Fullprof refinements.}
\label{MS_Figure-1}
\end{figure}

\begin{figure}
\centering \includegraphics [width = 0.48 \textwidth] {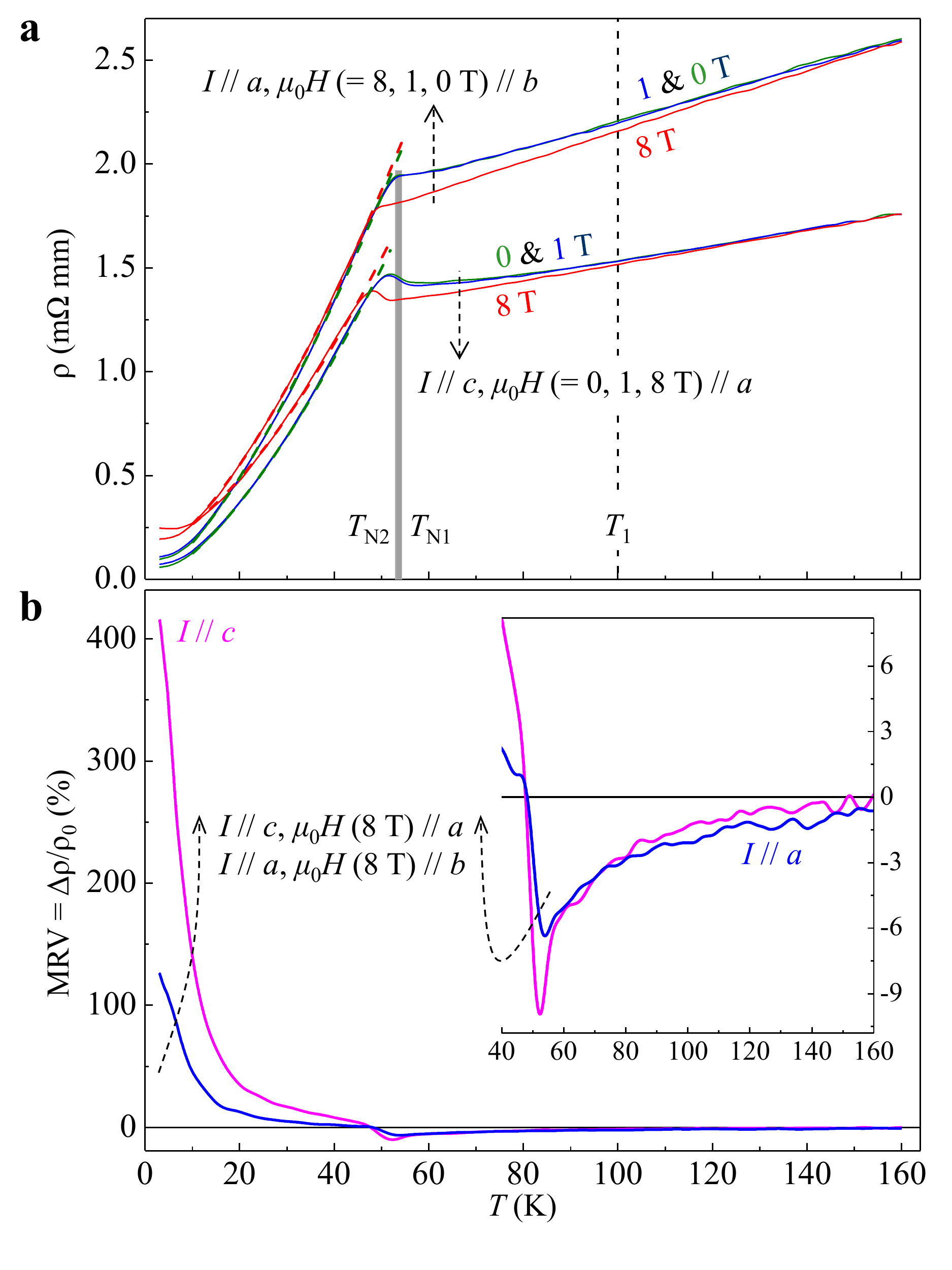}
\caption{\textbf{Temperature variations of resistivity and MR effect.}
(a) Resistivity measurements with current \emph{I} along the \emph{a} and \emph{c} axes under applied magnetic fields of 0, 1 and 8 T. $T_1$ = 100 K labels the temperature where one structural anomaly occurs as shown in Fig. 1. The dashed lines are fits between 10 K and 40 K (details in text), and extrapolated to higher temperatures. (b) Corresponding MR values versus temperature. The MR effect along the \emph{a} axis has a similar trend to that of the \emph{c} axis albeit with a lower value. The positive MR values decrease sharply with increasing temperature below $\sim$ 20 K, then gradually transfer into negative around $T_\texttt{N}$, and persist up to $\sim$ 120 K. By contrast, in amorphous Gd$_x$Si$_{1-x}$ films, only the NMRV was observed \cite{Castilho1991, Hellman1996, Hellman2000} near the metal-insulator transition. In (a), there is no big difference for the data at 0 and 1 T above $\sim$ 20 K. The solid lines in (a) and (b) in the dashed arrow direction in turn correspond to the ordinal axis-direction and applied magnetic-field strength as labeled, respectively.}
\label{MS_Figure-2}
\end{figure}

\begin{figure*}
\centering \includegraphics[width = 0.92 \textwidth] {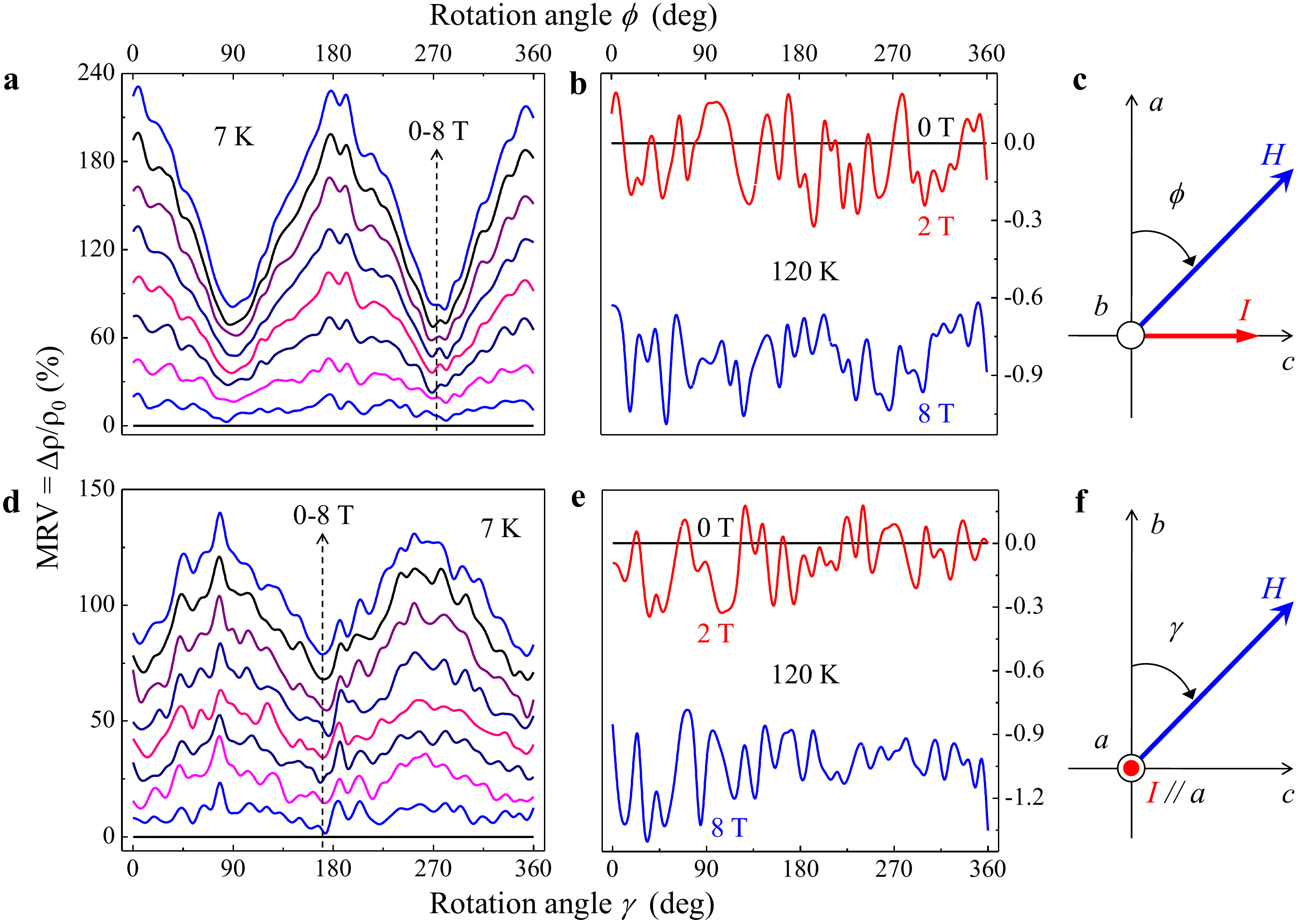}
\caption{\textbf{Angular-dependent MR values under different applied magnetic fields.}
(a) Angular-dependent MR values with applied magnetic field in a range of 0 to 8 T (1 T step) at 7 K, and (b) at 120 K (0, 2 and 8 T). (c) For (a) and (b) measurements, current \emph{I} is along the crystallographic \emph{c} axis, and applied magnetic field, $\mu_0$\emph{H}, rotates away from the \emph{a} axis with an angle of $\phi$ in the \emph{ac} plane. (d) Angular-dependent MR values with applied magnetic field in a range of 0 to 8 T (1 T step) at 7 K, and (e) at 120 K (0, 2 and 8 T). (f) For (d) and (e) measurements, \emph{I}$\|$\emph{a}-axis and applied magnetic field, $\mu_0$\emph{H}, rotates in the \emph{bc} plane with an angle of $\gamma$ deviated from the \emph{b} axis.}
\label{MS_Figure-3}
\end{figure*}

\begin{figure*}
\centering \includegraphics[width = 0.92 \textwidth] {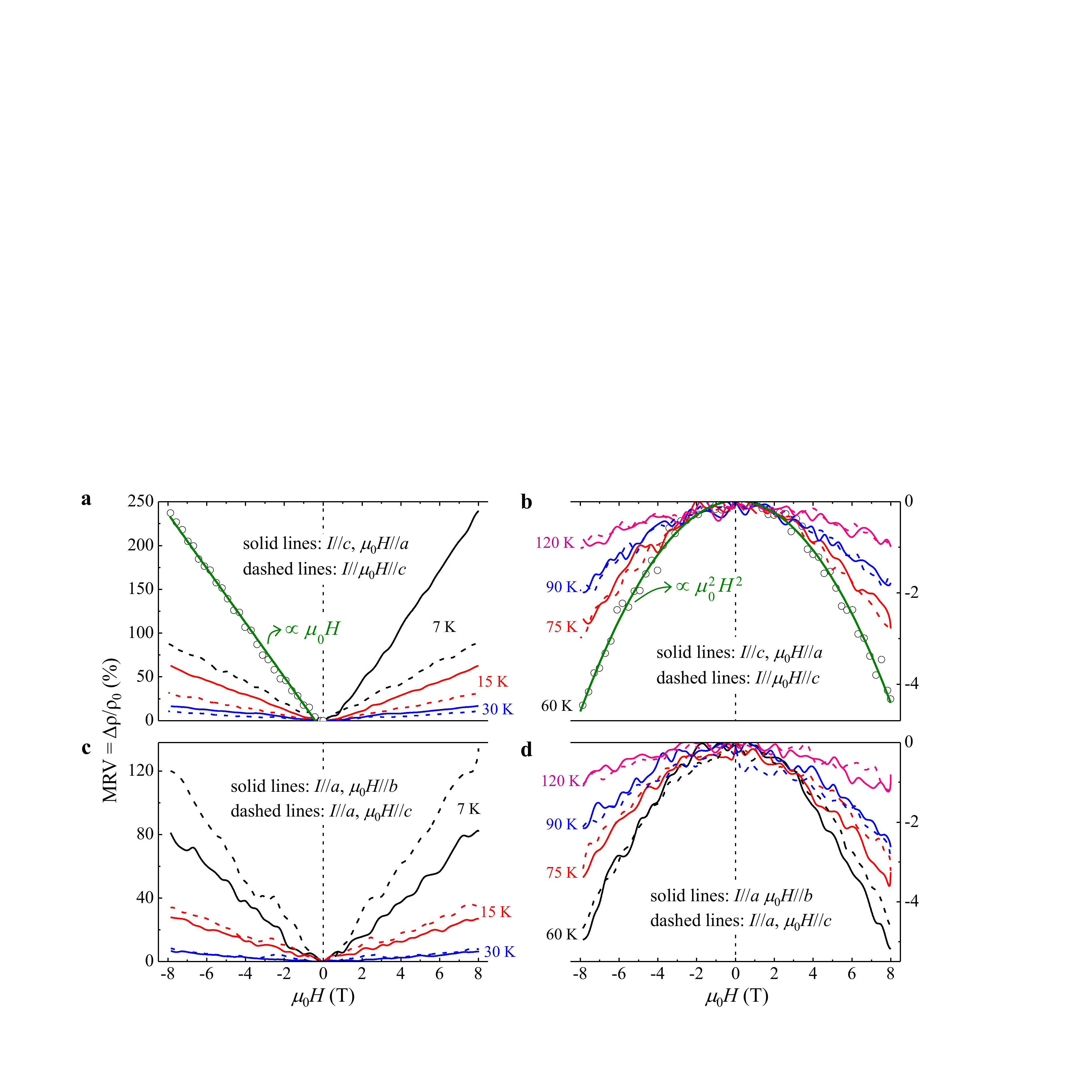}
\caption{\textbf{Field- and temperature- dependent MR values.}
(a) Field and temperature dependencies of the MR values with \emph{I}$\|$\emph{c}-axis and applied magnetic field, $\mu_0$\emph{H}, along the \emph{a} or \emph{c} axis below $T_\texttt{N}$, and (b) above $T_\texttt{N}$. The representatives of the linear-field dependence of the PMRV (i.e., the PMRV is proportional to the strength of applied magnetic field) at 7 K (blow $T_\texttt{N}$) and the quadratic variation of the NMRV (i.e., the absolute NMRV is proportional to the square of the strength of applied magnetic field) at 60 K (above $T_\texttt{N}$) were shown in (\texttt{a}) and (\texttt{b}), respectively. (c) Field and temperature dependencies of the MR values with \emph{I}$\|$\emph{a}-axis and applied magnetic field, $\mu_0$\emph{H}, along the \emph{b} or \emph{c} axis below $T_\texttt{N}$, and (d) above $T_\texttt{N}$.}
\label{MS_Figure-4}
\end{figure*}

\begin{figure*}
\centering \includegraphics[width = 0.92 \textwidth] {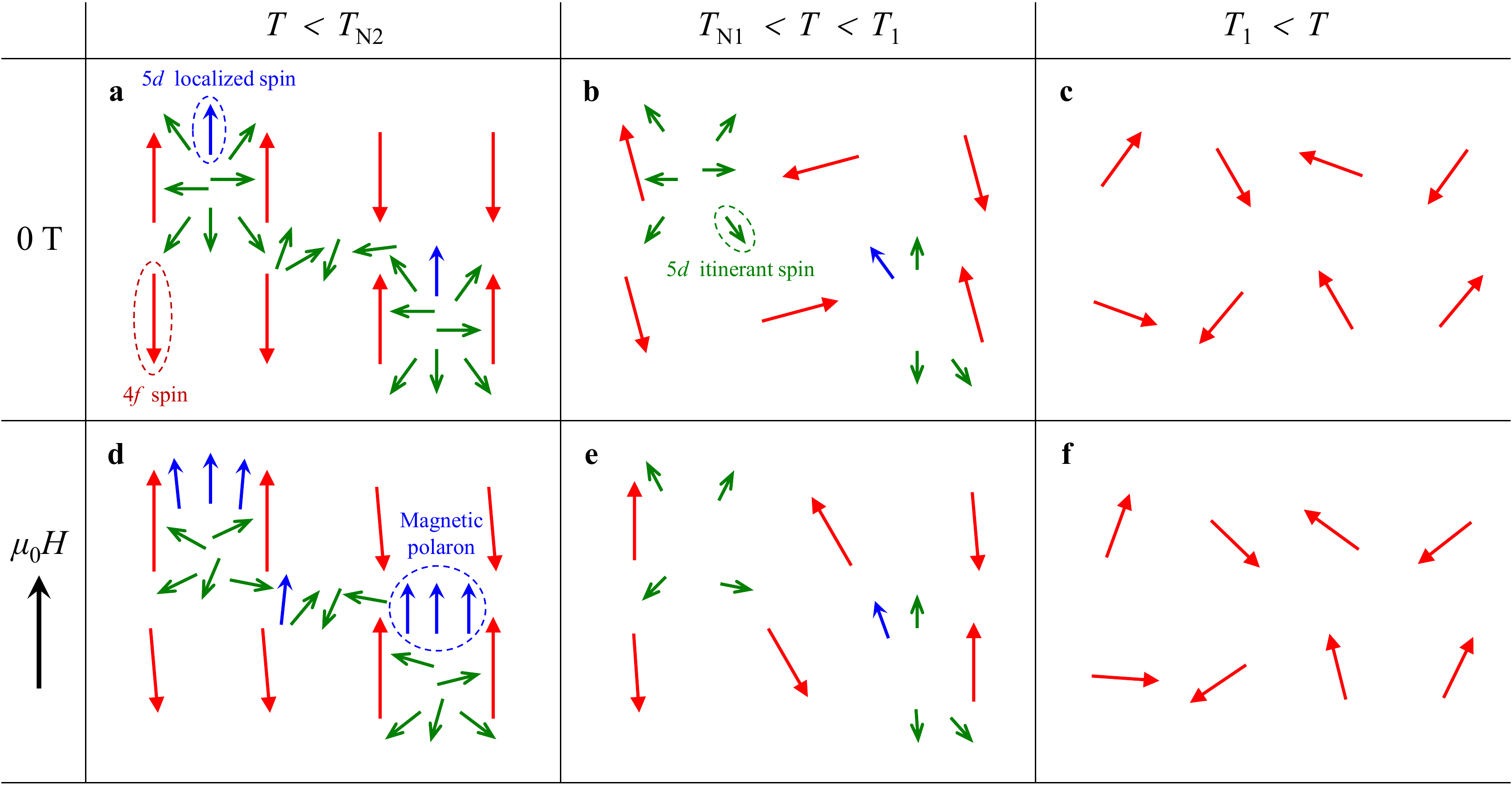}
\caption{\textbf{Schematic illustration of the spin states with and without applied magnetic field, $\mu_0$\emph{H}, in different temperature regimes.}
(a-c) At zero magnetic field. (d-f) At applied magnetic field of $\mu_0H$. When $T > T_1$, spin moments more or less rotate (f), depending on the strength of $\mu_0H$ and the size of MA, from a pure PA state (that is strictly observing the Curie-Weiss law as shown in Supplementary Fig. S4) in (c). When $T_{\texttt{N1}} < T < T_1$, the short-range AFM spins that are attributed only to the local 4\emph{f} moments appear (b) accompanied by the generations of polarized itinerate 5\emph{d} spins, based on the deviation of the unit-cell volume from the Gr$\ddot{\texttt{u}}$neisen model shown in Fig. 1b, and possible small amount of localized 5\emph{d} spins according to equation (1). Applied magnetic field mainly aligns the local AFM spins (e), leading to a decrease of the electron-spin scattering and resultant the NMRV. When $T < T_{\texttt{N2}}$, the LRO AFM state with almost equivalent AFM and FM interactions (see Supplementary Fig. S4) forms (a) with more itinerate 5\emph{d} moments (based on the large PMV effect shown in Fig. 1b). Applied magnetic field mainly localizes more 5\emph{d} moments by enhancing the exchange of \emph{J }in equation (1), resulting in the formation of magnetic polarons and the consequent PMRV (d).}
\label{MS_Figure-5}
\end{figure*}

\textbf{Results}

\textbf{Structural studies.} The x-ray powder diffraction analysis (Fig. 1) clearly shows three distinct structural regimes (I, II and III). Upon cooling, the refined (Re) \emph{a}, \emph{b}, \emph{c} and \emph{V} shrink almost linearly before $T_1$, followed by a slower decrease until sharp turns at $T_\texttt{N1}$, an onset temperature of the AFM transition. Below $T_\texttt{N1}$, the increases of \emph{a} and \emph{b} and the decrease of \emph{c} ultimately result in an unusual NTVE (i.e., PMV) in the unit-cell volume \emph{V}, which is quite useful in fabricating applicable materials with controlled thermal expansion values \cite{Barrera2005, Ibarra1995}. Obviously broadening in the nuclear Bragg peaks is present in the I regime (see Supplementary Fig. S1b), which is attributed to the magnetoelastic effect (the coupling between magnetic moments and lattice strains). The strain distribution patterns are extracted and shown in Supplementary Fig. S2. The variations in \emph{a}, \emph{b}, \emph{c} and \emph{V} below $T_\texttt{N1}$ imply that magnetic anisotropy (MA) (which is consistent with Supplementary Fig. S3a and uncommon for the \emph{S}-state Gd-compounds), spontaneous PMV and anisotropic MS effects exist in GdSi.

The CEF is mainly responsible for the giant MS effect in rare-earth (RE) compounds. This effect is thus expected to be negligible in GdSi. However, below $T_1$, structural parameters shown in Fig. 1a obviously deviate from the the
oretical estimates (solid lines) by the Gr$\ddot{\texttt{u}}$neisen (Gr) law, e.g., $\frac{a^\texttt{10K}_\texttt{Re} - a^\texttt{10K}_\texttt{Gr}}{a^\texttt{10K}_\texttt{Gr}}$ = 4.94(6) $\times$ 10$^{-4}$, $\frac{b^\texttt{10K}_\texttt{Re} - b^\texttt{10K}_\texttt{Gr}}{b^\texttt{10K}_\texttt{Gr}}$ = 8.24(9) $\times$ 10$^{-4}$ and
$\frac{c^\texttt{10K}_\texttt{Re} - c^\texttt{10K}_\texttt{Gr}}{c^\texttt{10K}_\texttt{Gr}}$ = -2.60(8) $\times$ 10$^{-4}$, denoting large anisotropic spontaneous MS effects. The formation of long-range-ordered (LRO) AFM state is a process of the growth of sublattice FM domains. The enlargement of FM domain volumes with decreasing temperature may accumulate strains on the domain walls, which is the microscopic mechanism for the magnetic-field-induced MS effect in ferromagnets. Therefore, including the effect of the molecular field of one Gd-AFM-sublattice on the other is indispensable to understand the
\emph{spontaneous} MS effect in GdSi. According to the Stoner model for itinerant magnetic electrons, the positive magnetic pressure $P_M$ associated with the magnetic ordering in a band is proportional to $\frac{\partial \texttt{ln}D}{\partial \texttt{ln}V}M^2$, where \emph{D}, \emph{V} and \emph{M} represent the electronic density of states at the Fermi energy, the volume and the magnetic moment, respectively. The spontaneous PMV effect (i.e., NTVE), e.g., $\frac{V^\texttt{10K}_\texttt{Re} - V^\texttt{10K}_\texttt{Gr}}{V^\texttt{10K}_\texttt{Gr}}$ = 1.04(2) $\times$ 10$^{-3}$, is therefore attributed mainly to the increases of \emph{D} (corresponding to the pronounced decrease of $\rho$ below $T_\texttt{N}$ in Fig. 2a) and the induced itinerate spin-moments in conduction bands. Similar MS and MV effects were also reported in Gd$_3$Ni \cite{Kusz2000}, where, however, they are ascribed to the itinerant character of the Ni 3\emph{d} bands.

\textbf{Resistivity measurements.} As depicted in Fig. 2a, the zero-field-cooling (ZFC) electrical resistivity at 0 T with current \emph{I} along the \emph{a} and \emph{c} axes decreases linearly due to the weakened thermal excitations upon cooling until around $T_\texttt{N}$ (vertical bar), where a hump with negative slop appears along only the \emph{c} axis, probably attributed to the magnetic superzone effect \cite{Mackintosh1962} as a consequence of the AFM ordering. Below $T_\texttt{N}$, they decrease steeply like a reasonable AFM metal. The resistivity between 10 K and 40 K \cite{Yamada1974} can be well fit to $\rho$(\emph{T}) = $\rho_0$ + \emph{k}\emph{T}$^w$, shown as dashed lines. This produces $w^{c, \texttt{0T}}$ = 1.51(2), $w^{c, \texttt{8T}}$ = 1.56(4), $w^{a, \texttt{0T}}$ = 1.43(2) and $w^{a, \texttt{8T}}$ = 1.51(1). All \emph{w} values are much smaller than 5 \cite{Yamada1974} indicative of anisotropic magnetic interactions.

\textbf{Anisotropic MR effect and positive MR value (PMRV) to NMRV transition.} The most intriguing results from resistivity measurements are the anisotropic MR effect (Figs 2b, 3 and 4) and the existence of both positive and negative MR values, up to $\sim$ 415\% (comparable to the CMR value in manganites \cite{Santen1950, Urushibara1995} and one to two orders of magnitude larger than that of the RE-metals \cite{Mackintosh1964}) and down to $\sim$ -10.5\% along the \emph{c} axis at 8 T and 3 K and 52.8 K, respectively. The MR anisotropy in the \emph{ac} and \emph{bc} plans is shown in Fig. 3. They display a twofold symmetry at 7 K (Figs 3a and 3d). We notice that applied magnetic field of 8 T does not suppress (produce) the (a) hump along the \emph{c} and \emph{a} axes, respectively, near $T_\texttt{N}$, and the MR twofold symmetry is persistent from 1 to 8 T, indicating that applied magnetic field in a strength of 8 T may just align or localize the 5\emph{d} moments, and slightly rotate and tilt the 4\emph{f} moments while conserving the superzone energy gap. Therefore, the MR effect in GdSi exhibits a well separate feature of the temperature regions for the positive and the negative MR values (Fig. 2b), respectively, which is induced jointly by the AFM superzone effect \cite{Mackintosh1962} and the shift of the AFM transition to lower temperatures in external applied magnetic field analogous to the case in manganites \cite{Santen1950, Jin1994, Urushibara1995}. For metals, mean-field theories predict that spin fluctuations induced by applied magnetic field from the localized magnetism produce a PMRV with the quadratic-field dependence in antiferromagnets, whereas a NMRV with the linear variation in ferromagnets and paramagnets \cite{Yamada1973}. However, in GdSi, the PMRV in the AFM state mainly displays a linear magnetic-field dependence (Figs 4a and 4c), while above $T_{\texttt{N1}}$ the absolute NMRV is proportional to the square of the strength of applied magnetic field (Figs 4b and 4d). Both the positive and negative MR effects do not saturate at utilized maximum $\mu_0$\emph{H} = 8 T (Fig. 4). In addition, the ratio of the resistivity at 160 K and 7 K in Fig. 2a is already $\sim$ 12-25, therefore, the cyclotron motion of the conduction electrons could be neglected at $\mu_0H$ = 8 T (i.e., $\omega_\texttt{c} \tau \ll 1$, where $\omega_\texttt{c}$ is the cyclotron frequency and $\tau$ is the life time of the conduction electrons). Therefore, these uncommon magnetic-field variations indicate new transport mechanisms for the MR effects of GdSi.

\textbf{Discussion}

In GdSi, the conduction electrons (mainly 5\emph{d} bands) are different from those responsible for the magnetism (4\emph{f} component plus possible part of the 5\emph{d} component). The former is normally delocalized, acting as the magnetic glue among magnetic ions (Fig. 5), and scattered by them, leading to electrical resistance. The magnetism from the 4\emph{f} part is generally localized with weak interactions. Therefore, the LRO AFM state originates mainly from the isotropic RKKY interactions through conduction bands \cite{Jensen1991}. The interaction between localized moments, $\overrightarrow{M}_{\texttt{loc}}$, and itinerant ones, $\overrightarrow{m}_{\texttt{iti}}$, can generate an extraordinarily large Zeeman splitting in the mean-field approximation \cite{Moln2007}
\begin{eqnarray}
E = g^\ast \mu_\texttt{B} \overrightarrow{m}_{\texttt{iti}} \cdot H + 2J(H) \overrightarrow{m}_{\texttt{iti}} \cdot \langle \overrightarrow{M}_{\texttt{loc}} \rangle,
\end{eqnarray}
where $g^\ast$ is the spectroscopic splitting factors for the carriers, $\mu_\texttt{B}$ is the Bohr magneton, \emph{J} is the effective exchange coefficient and $\langle \overrightarrow{M}_{\texttt{loc}} \rangle$ is the averaged local moment in the regime of band electrons. Since $\overrightarrow{m}_{\texttt{iti}}$ is usually small, and the second term could be very large (e.g., amounting to fractions of an eV in the LRO magnetic state of Eu-compounds \cite{Moln2007}), in addition, \emph{J} is strongly associated with applied magnetic field by virtue of modifying spin fluctuations of $\overrightarrow{M}_{\texttt{loc}}$, the formation of magnetic polarons in the 5\emph{d} bands by this splitting in the LRO AFM state of GdSi is thus possible. Therefore, when $T < T_{\texttt{N2}}$, the modified \emph{J} at 8 T drives some of the conducting moments (as foregoing remarks) to form local magnetic polarons that lead to a largely degenerate conduction (i.e., PMRV) (Figs 5a and 5b). In this case, the more extended 5\emph{d} bands almost certainly offer a small FM component, which dominates the linear-magnetic-field dependence below $T_{\texttt{N2}}$.

Above $T_{\texttt{N1}}$, the LRO AFM 4\emph{f} moments disappear, which is implied by the change of the $\chi$ relationship between the \emph{a}, \emph{b} and \emph{c} axes (see Supplementary Fig. S3a). However, the upward deviation of 1/$\chi$ from the Curie-Weiss law below $T_1$ (see Supplementary Fig. S4) along with the concomitant anomalous volume lattice distortions (Fig. 1) indicates that a kind of striking AFM state exists there, which can be ascribed only to the local short-range AFM 4\emph{f} moments (Fig. 5b) according to equation (1) (broad 5\emph{d} states indeed could not form a magnetic state by themselves). This local AFM state theoretically \cite{Yamada1973} leads to the quadratic-magnetic-field variation (Figs 4b and 4d). The local AFM moments above $T_{\texttt{N1}}$ are possibly oriented irregularly in the PM state and become compulsorily aligned and to some extent spin-tilted (Fig. 5e), leading to the decrease in resistivity and thereby promoting a NMRV.

Including the contribution of magnetic polarons into the mean-field approximation of the MR effect of AFM metals \cite{Yamada1973, Yosida1957, Bogach2006, Anisimov2009}, the MR value (MRV) then equals
\begin{eqnarray}
|\texttt{MRV}| & \approx & A |\overrightarrow{M}^{\texttt{4f}}_{\texttt{In}} + \overrightarrow{M}^{\texttt{5d}}_{\texttt{In}}|^2 \nonumber \\*
& \approx & A (|\overrightarrow{M}^{\texttt{4f}}_{\texttt{In}}|^2 + 2 |\overrightarrow{M}^{\texttt{4f}}_{\texttt{In}}| |\overrightarrow{M}^{\texttt{5d}}_{\texttt{In}}| \cos\alpha + |\overrightarrow{M}^{\texttt{5d}}_{\texttt{In}}|^2),
\end{eqnarray}
where \emph{A} is a constant, $M_{\texttt{In}}$ is the induced local magnetization by applied magnetic field, $\mu_0H$, $\alpha$ is the angle between $\overrightarrow{M}^{\texttt{4f}}_{\texttt{In}}$ and $\overrightarrow{M}^{\texttt{5d}}_{\texttt{In}}$, $\overrightarrow{M}^{\texttt{4f}}_{\texttt{In}} = \overrightarrow{\chi}^{\texttt{4f}}_{\texttt{In}}\mu_0 H$, and $\overrightarrow{\chi}^{\texttt{4f}}_{\texttt{In}}$ is the magnetic susceptibility of the 4\emph{f} site along the AFM sublattice direction. At $T_{\texttt{N1}} < T < T_1$, neglecting $|\overrightarrow{M}^{\texttt{5d}}_{\texttt{In}}|$, $|\texttt{NMRV}| \propto |\overrightarrow{M}^{\texttt{4f}}_{\texttt{In}}|^2 = |\overrightarrow{\chi}^{\texttt{4f}}_{\texttt{In}}|^2 \mu_0^2 H^2$, resulting in a quadric-magnetic-field dependence. At $T < T_{\texttt{N2}}$, in our case, the linear part in equation (2) is dominate, i.e., PMRV $\propto |\overrightarrow{M}^{\texttt{4f}}_{\texttt{In}}| |\overrightarrow{M}^{\texttt{5d}}_{\texttt{In}}| \cos\alpha = |\overrightarrow{\chi}^{\texttt{4f}}_{\texttt{In}}| |\overrightarrow{M}^{\texttt{5d}}_{\texttt{In}}| \mu_0H \cos\alpha$, which produces not only the linear-magnetic-field dependence but also the twofold MR symmetry (Figs 3a and 3d). Quantitative analysis requires the knowledge of the exact directions of $\overrightarrow{\chi}^{\texttt{4f}}_{\texttt{In}}$ and $\overrightarrow{M}^{\texttt{5d}}_{\texttt{In}}$. The involvement of magnetic polarons in understanding the MR effects here is further supported by the existence of strong FM Gd-Gd interactions (see Supplementary Fig. S4).

RE 4\emph{f} electrons generally remain localized so that their properties in an alloy closely resemble those in the single-elemental metals. It is therefore interesting to compare the microscopic conducting mechanism of GdSi with that of the 3\emph{d} transition metals \cite{Mott1936, Wilson1938}. Roughly, the 4\emph{f} shells highly screened by the 5\emph{s} and 5\emph{p} states are better shielded than the transition metal (TM) 3\emph{d} ones. Therefore, the 4\emph{f} electrons are well embedded within the atom and the 5\emph{d} and 6\emph{s} states act as conduction electrons in the metals, though band structure calculations suggested that the upper part of the 4\emph{f} spin minority may hybridize with the unoccupied 5\emph{d} and 6\emph{s} bands \cite{Singh1991}. On the contrary, for 3\emph{d} TMs, the conductivity are mainly dominated by the \emph{s}-\emph{s} and \emph{d}-\emph{d} transitions at low temperatures, in addition to the \emph{s}-\emph{d} transitions (increasing the effective electron mass and shortening the mean free path) at high temperatures \cite{Mott1936, Wilson1938}, supposing that both \emph{s} and \emph{d} electrons are conduction electrons. Since almost all RE 4\emph{f} metals and only some of the 3\emph{d} TMs (Fe, Co, Ni and Mn) order magnetically at low temperatures, magnetism plays a more crucial role in controlling electrical property of RE 4\emph{f} metals.

In summary, GdSi displays a wide array of novel behaviors including a two-step AFM transition, an anisotropic spontaneous MS effect, magnetostrictive strains, in particular, an exotic NTVE (spontaneous PMV) and a peculiar positive to negative MR transition, indicating a strong coupling between spin, lattice and charge degrees of freedom. The peculiar PMRV to NMRV transition and their nontrivial \cite{Yamada1973} magnetic-field dependencies are predominated by the nature of the 4\emph{f} moments and the polaronic 5\emph{d} carriers. Coexistence of these extraordinary behaviors in the same Gd-compound is unique. The present results make GdSi an fascinating system for theoretical and further experimental explorations.

\textbf{Methods}

Since the neutral Gd is a strong thermal neutron absorber, we therefore performed a powder x-ray diffraction study of the pulverized GdSi single crystal on an in-house diffractometer employing the copper $K_{\alpha 1}$ = 1.5406(9) $\texttt{\AA}$ as the radiation with a 2$\theta$ step size of 0.005$^\circ$ from 10 to 300 K to explore the temperature-dependent structural modifications. X-ray powder diffraction data were analyzed by Fullprof suite \cite{Rodriguez-Carvajal1993}. ZFC dc magnetization (\emph{M}) measurements were performed on a Quantum Design MPMS-7 SC quantum interference device (SQUID) magnetometer. The ZFC electrical resistivity ($\rho$) of bar-shaped ($\sim$ 0.6 $\times$ 1 $\times$ 6 mm$^3$) single crystals was measured by standard DC four-probe technique using a commercial physical property measurement system (PPMS), equipped with the option of rotating the sample \emph{in situ}. Commercial silver paint and 25 $\mu$m gold wire were used for electrical contacts. MRV($\mu_0H,T,\phi/\gamma) = \frac{\rho(\mu_0H,T,\phi/\gamma)-\rho(0,T,\phi/\gamma)}{\rho(0,T,\phi/\gamma)} \times 100\%$, where $\rho(\mu_0H, T, \phi/\gamma)$ and $\rho(0, T, \phi/\gamma)$ are the resistivity with and without $\mu_0$\emph{H} at a given temperature, \emph{T}, and a rotating angle of $\phi$ or $\gamma$ (Figs 3c and 3f).

\textbf{Acknowledgements}

H.F.L thanks M-T Fernandez-Diaz at Institut Laue-Langevin, France, for helpful discussions, and is grateful to E. Kentzinger at Forschungszentrum J$\ddot{\texttt{u}}$lich GmbH, Germany, for his efforts in keeping the high performance of the x-ray powder diffractometer. This work at RWTH Aachen University and J$\ddot{\texttt{u}}$lich Centre for Neutron Science JCNS Outstation at Institut Laue-Langevin was funded by the BMBF under contract No. 05K10PA3.

\textbf{Author contributions}

H.F.L, Y.X, B.S, J.P and P.M characterized the samples by Laue backscattering, temperature-dependent x-ray powder diffraction, SQUID, PPMS, etc. H.F.L, Y.X, W.S, G.R, and Th.B discussed and analyzed the results. H.F.L led the project and wrote the paper.

\textbf{Additional information}

\textbf{Supplementary information} 

\textbf{Competing financial interests:} The authors declare no competing financial interests.

\textbf{License:} 

\textbf{How to cite this article:} 

\newpage
\appendix

\section{Supplementary information for \textquotedblleft Possible magnetic-polaron-switched positive and negative magnetoresistance
in the GdSi single crystal\textquotedblright}

\textbf{X-ray powder diffraction refinements and anisotropic stains.} The observed and refined x-ray powder patterns at 10 K and 300 K are shown in Fig. S1a. The listed reliability factors in the caption indicate good refinements. In Fig. S1b, the full width at half maximum of the nuclear (1 1 2) Bragg peak by Lorentzian fitting increases largely below $T_\texttt{N}$, indicating the existence of strong strains that are produced by the gradual formation of the antiferromagnetic (AFM) state that is a process accompanied by the AFM sublattice domain movements. We first measured the standard LaB$_6$ powders, with which we extracted the instrumental resolution. Then we refined the collected x-ray powder patterns at 10 K with the obtained resolution using Fullprof suite \cite{Rodriguez-Carvajal1993} to extract the stain configurations as shown in Fig. S2. The existing AFM-driven anisotropic strains conserve the lattice symmetry or may lead to possible nuclear structural domain changes, which, however, may be too small to be detected by the instrument accuracy. It is stressed that the lattice change in GdSi, e.g., $\frac{c_{\texttt{54.8K}} - c_{\texttt{10K}}}{c_{\texttt{54.8K}}}$ = 3.2(1) $\times$ 10$^{-4}$, is much smaller than that of the symmetry-broken AFM metals such as SrFe$_2$As$_2$, where $\frac{b_{\texttt{202K}} - b_{\texttt{20K}}}{b_{\texttt{202K}}}$ = 3.1(2) $\times$ 10$^{-3}$ \cite{Haifeng2009}. Similar AFM-driven strains were also observed in the classical 3\emph{d} Chromium metal that also displays charge- and AFM-spin-density waves \cite{Fawcett1988}.

\textbf{Magnetization measurements, AFM transitions and magnetic anisotropy.} To confirm that the positive to negative magnetoresistance (MR) transition is really induced by the shift of $T_\texttt{N}$ to low temperatures with applied magnetic fields, we exactly fixed the crystallographic \emph{a} and \emph{c} axes, taking advantage of the rotator, and measured the temperature-dependent magnetization as shown in Fig. S3. It is unexpected that the AFM transition experiences two steps at $T_{\texttt{N1}}$ and $T_{\texttt{N2}}$ (Fig. S3a), respectively. From $T_{\texttt{N1}}$ to $T_{\texttt{N2}}$, there is a slight decrease in magnetization, whereas below $T_{\texttt{N2}}$ it first-order-like diminishes rapidly indicative of the formation of the long-range-ordered AFM state. Note that $\chi_b > \chi_c > \chi_a$ below $T_{\texttt{N2}}$ transfers to $\chi_c > \chi_b > \chi_a$ between $T_{\texttt{N2}}$ and $T_{\texttt{N1}}$, and subsequent to $\chi_c > \chi_a > \chi_b$ above $T_{\texttt{N1}}$, indicating the existence of magnetic anisotropy (MA) and different magnetic states in the three temperature regimes. The observed MA here is in agreement with the anomalous lattice constants (Fig. 1a), the lower \emph{w} values than 5 (Fig. 2a), and the approximately zero paramagnetic Curie temperatures as discussed below (Fig. S4). Applied magnetic fields indeed shift $T_{\texttt{N1}}$ and $T_{\texttt{N2}}$ to lower temperatures in the constructed $\mu_0$\emph{H}-\emph{T} phase diagram shown in Fig. S3b indicative of some local Gd-Gd AFM interactions, which is dramatically different with the purely itinerant Cr whose $T_\texttt{N}$ is strongly rigid in response to applied magnetic fields even up to 14 T, nevertheless extremely sensitive to strains \cite{Fawcett1988}.

\begin{figure*} \centering \includegraphics [width = 0.94\textwidth] {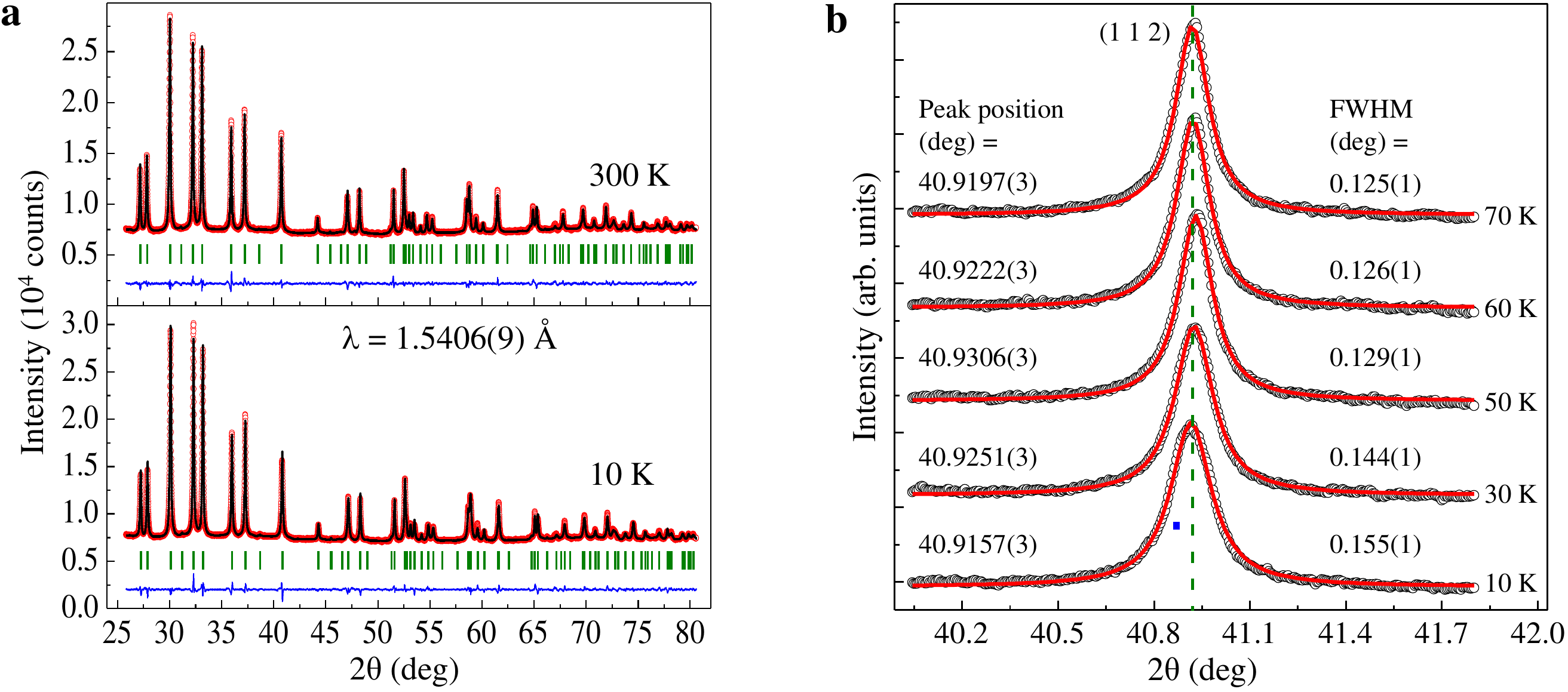}
\caption{\textbf{Powder x-ray diffraction patterns and refinements.}
\textbf{a,} Observed (circles) and calculated (solid lines) x-ray powder-diffraction patterns at 10 K (down) and 300 K (up). The vertical bars mark the positions of Bragg reflections. The lower curve represents the difference between the observed and calculated patterns. The reliability factors of refinements are: 10 K $[R_p (\%) = 1.06, R_{wp} (\%) = 1.57, \chi^2 = 1.37, R_B (\%) = 2.65, R_F (\%) = 2.84]$; 300 K $[R_p (\%) = 0.961, R_{wp} (\%) = 1.33, \chi^2 = 1.22, R_B (\%) = 1.47, R_F (\%) = 1.70]$. \textbf{b,} Observed nuclear (1 1 2) Bragg peaks at selected temperatures from 10 K to 70 K (below $T_\texttt{N2}$ and above $T_\texttt{N1}$). The solid lines are fits of the Lorentzian line-shape, and the dashed line is a guide to the eye. The small horizontal bar indicates the instrumental resolution. The fitted peak position and full width at half maximum are labeled at respective temperatures.}
\label{SUP_Figure-1}
\end{figure*}

\begin{figure*}
\centering \includegraphics[width = 0.94\textwidth] {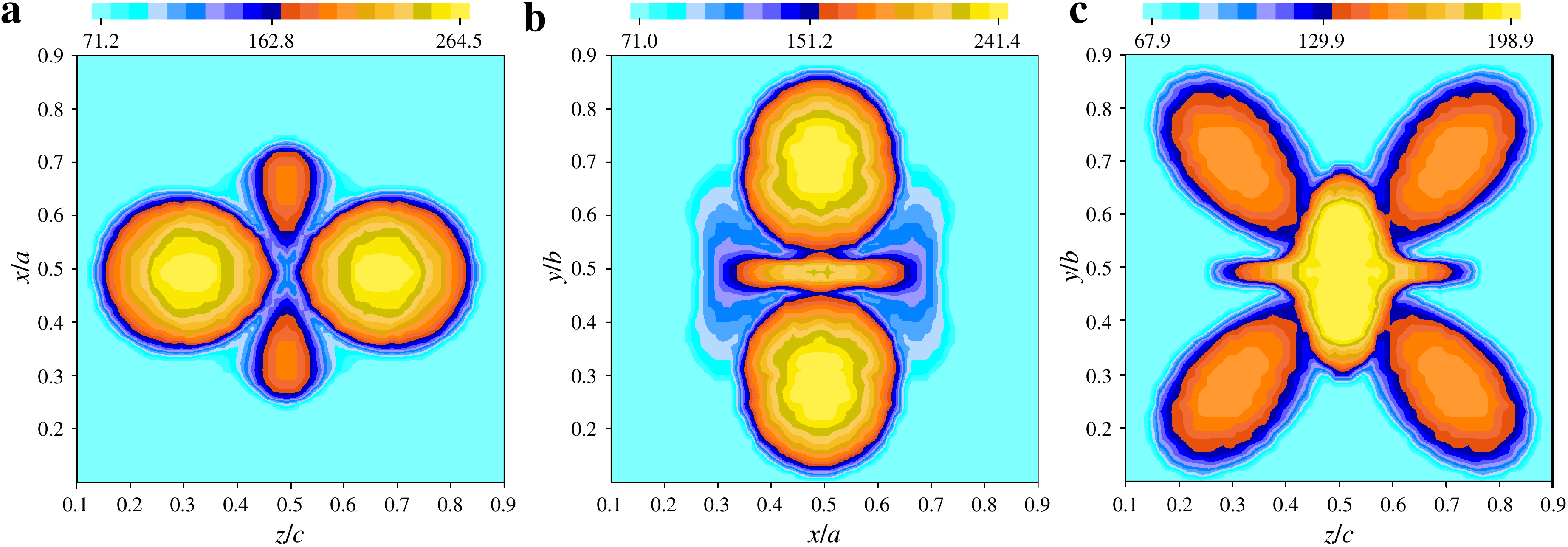}
\caption{\textbf{Strain patterns of GdSi.}
Extracted anisotropic stain configurations in the \emph{ac} (\textbf{a}), \emph{ba} (\textbf{b}) and \emph{bc} (\textbf{c}) planes at 10 K.}
\label{SUP_Figure-2}
\end{figure*}

\begin{figure*} \centering \includegraphics [width = 0.94\textwidth] {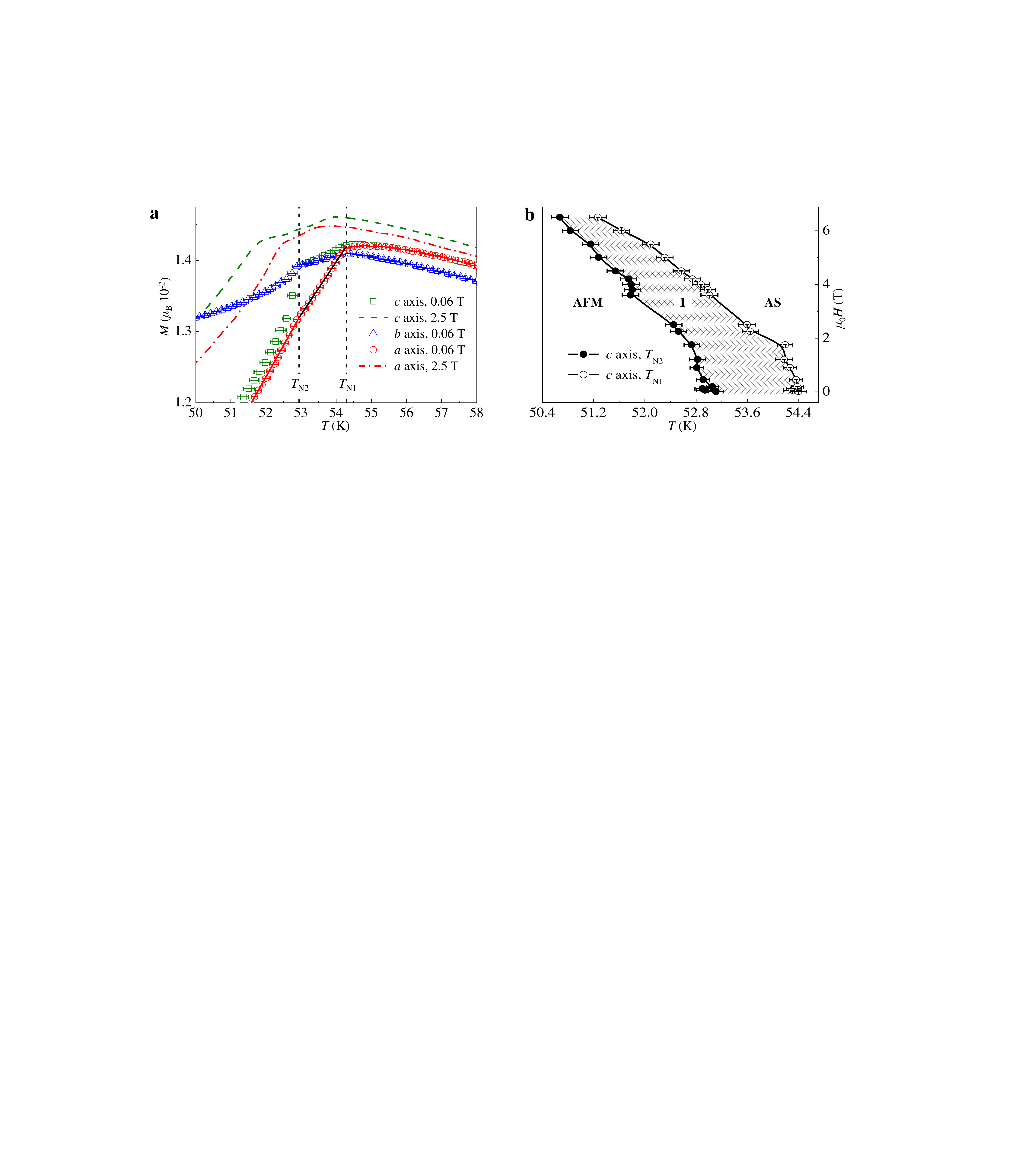}
\caption{\textbf{Antiferromagnetic transitions and $\mu_0$\emph{H}-\emph{T} phase diagram.}
\textbf{a,} Temperature-dependent magnetization under $\mu_0$\emph{H} = 0.06 T (along the \emph{a}, \emph{b} and \emph{c} axes) and 2.5 T (along the \emph{a} and \emph{c} axes, shifted down for comparison). The solid lines are linear fits of the \emph{a}-axis magnetization (0.06 T) between $T_{\texttt{N2}}$ ($\sim$ 52.94 K) and $T_\texttt{\texttt{N1}}$ ($\sim$ 54.3 K), and just below $T_{\texttt{N2}}$ (50-52.94 K), indicating a clear slop change from 7.4(1) $\times$ 10$^{-6}$ (above $T_{\texttt{N2}}$) to 8.8(1) $\times$ 10$^{-6}$ $\mu_\texttt{B}$/K (below $T_{\texttt{N2}}$). \textbf{b,} Measured $\mu_0$\emph{H}-\emph{T} phase diagram around $T_{\texttt{N1}}$ and $T_\texttt{N2}$. Error bars are the standard deviation based on entire measurements. The solids are guides to the eye. The intermediate (I) phase bordering the long-range-ordered static 4\emph{f} AFM state and the local short-range 4\emph{f} AFM spins (AS) may be attributed to a short-range-ordered 4\emph{f} AFM state because of the large suppression of $T_{\texttt{N1}}$ and $T_{\texttt{N2}}$, consistent with the stiffness of the hump in resistivity under applied magnetic fields (Fig. 2a).} \label{SUP_Figure-3}
\end{figure*}

\begin{figure*} \centering \includegraphics [width = 0.94\textwidth] {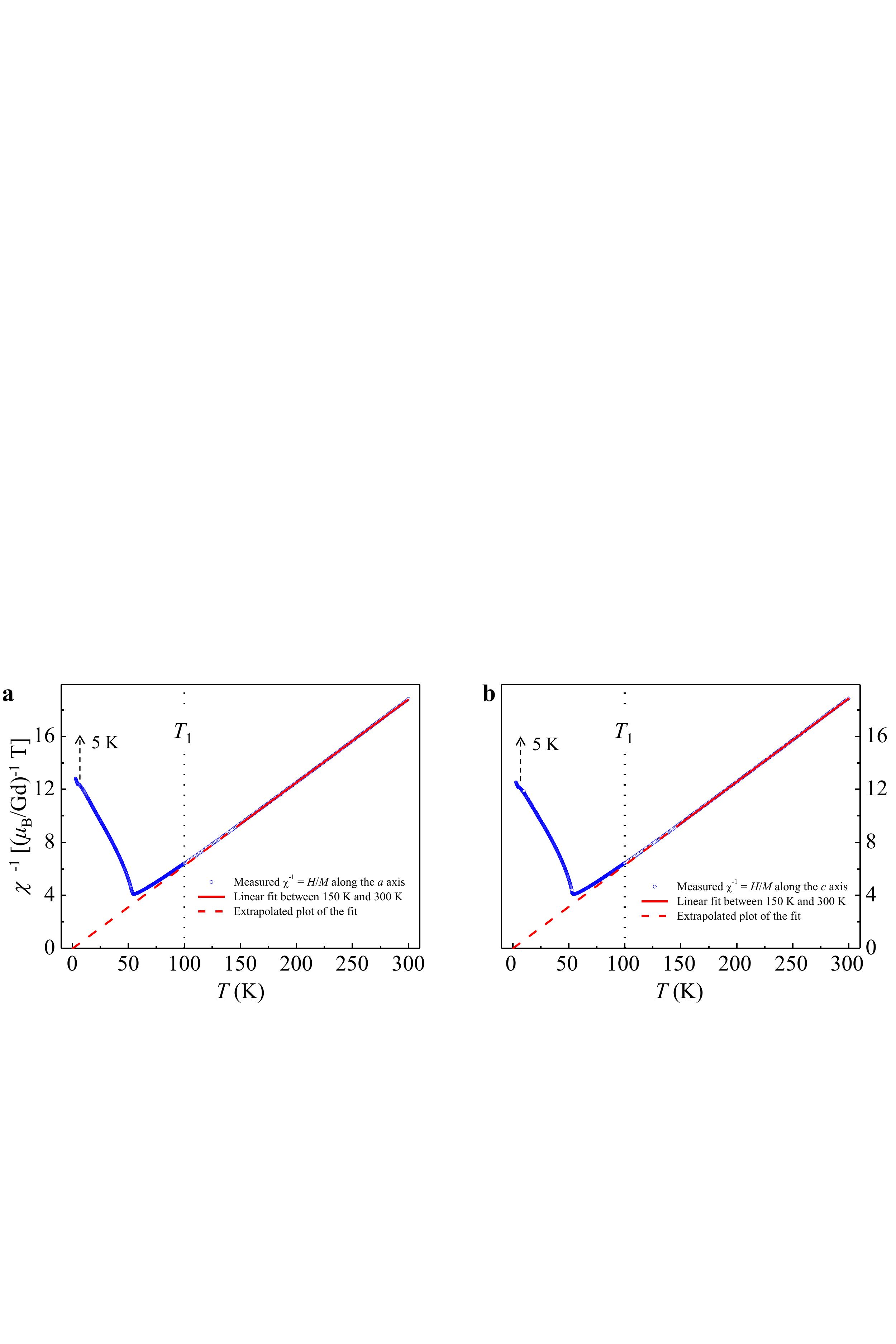}
\caption{\textbf{Inverse susceptibility versus temperature.} As measured (circles) at 0.06 T along the \emph{a} (\textbf{a}) and \emph{c} (\textbf{b}) axes in GdSi. The solid lines indicate a Curie-Weiss behavior between 150 K and 300 K, which was extrapolated to zero temperature (dashed lines). One anomaly displays at $\sim$ 5 K. $T_1$ $=$ 100 K labels the temperature where one structural anomaly occurs as shown in Fig. 1a.}
\label{SUP_Figure-4}
\end{figure*}

We also measured the temperature dependence of the magnetic susceptibility $\chi$ = \emph{M}/$\mu_0$\emph{H} along the \emph{a} and \emph{c} axes at 0.06 T in a temperature range of 3-300 K. The corresponding inverse susceptibility 1/$\chi$ is shown in Fig. S4. Fitting the data between 150 K and 300 K to the Curie-Weiss (CW) law, we obtained the effective magnetic moment, $\mu_{\texttt{eff}}$, and the paramagnetic Curie temperature, $\theta_\texttt{P}$, to be $\mu^{[100]}_{\texttt{eff}}$ = 8.437(2) $\mu_{\texttt{B}}$/Gd and $\theta^{[100]}_\texttt{P}$ = 0.33(1) K for $\mu_0$\emph{H} along the [100] direction, and $\mu^{[001]}_{\texttt{eff}}$ = 8.424(1) $\mu_{\texttt{B}}$/Gd and $\theta^{[001]}_\texttt{P}$ = 0.09(7) K for $\mu_0$\emph{H} parallel to the [001] direction. Below $\sim$ 100 K, $\chi^{-1}(T)$ upward deviates largely from the CW law, implying an increasing contribution of the \emph{AFM} spin correlations. It is pointed out that small upward deviation already appears at $\sim$ 120 K consistent with the observed negative MR effect at 120 K as shown in Figs 3b, 3e, 4b and 4d. However, this kind of small local AFM spin states do not generate appreciable anomalies in the structural parameters until upon cooling to a temperature of $\sim$ 100 K, where a sufficiently large magnetoelastic coupling that is able to shake the crystalline lattice may be just attained, as shown in Fig. 1. The occurrence of very small positive paramagnetic curie temperatures, $\theta_\texttt{P}$, in the dominate AFM matrix, indicates that either the 4\emph{f} moments in GdSi are purely localized without any interaction at all, or there exist mixed ferromagnetic (FM) and AFM Gd-Gd interactions with an approximatively counterbalance of their strengths like the rare case in AFM ZnCr$_2$S$_4$ \cite{Hemberger2006}. The coexistence of FM and AFM exchange interactions often causes spin frustrations.

\textbf{Magnetic anomaly at $\sim$ 5 K.} The anomaly at $\sim$ 5 K in Fig. S4 may suggest a different magnetic state below 5 K with the one between 5 K and $T_{\texttt{N2}}$. A similar anomaly was also observed in Ref. \cite{Tung2005}. The fitting of the resistivity between 10 K and 40 K (Fig. 2a) also produces $\rho^{c, \texttt{0T}}_0$ = -7.3(2) $\times$ 10$^{-3}$ m$\Omega$ mm, $\rho^{c, \texttt{8T}}_0$ = 1.5(1) $\times$ 10$^{-1}$ m$\Omega$ mm, $\rho^{a, \texttt{0T}}_0$ = -1.0(2) $\times$ 10$^{-2}$ m$\Omega$ mm and $\rho^{a, \texttt{8T}}_0$ = 1.1(1) $\times$ 10$^{-1}$ m$\Omega$ mm. It is interesting that both $\rho^{c, \texttt{0T}}_0$ and $\rho^{a, \texttt{0T}}_0$ (at zero K) are negative, indicating that at lower temperatures the conduction mode may be modified \cite{Yamada1974}, or there exist magnetic transitions consistent with the observed clear anomalies at $\sim$ 5 K in magnetization measurements (Fig. S4).

We did not observe any anomaly indicative of the weak FM order at $\sim$ 78 K shown in Ref. \cite{Roger2006}.

\textbf{Future studies.} Determining the magnetic structures and interactions will be the subjects for future studies, which may be not easy because of the strong thermal neutron absorption of neutral Gd and the intangible influence of 5\emph{d} conducting moments on the 4\emph{f} AFM ones.


\begin{thebibliography}{99}
\bibitem{Santen1950} Van Santen, J. H. \& Jonker, G. H. Electrical conductivity of ferromagnetic compounds of manganese with perovskite structure.
    \emph{Physica}, \textbf{16}, 599-600 (1950).
\bibitem{Jin1994} Jin, S., Tiefel, T. H., McCormack, M., Fastnacht, R. A., Ramesh, R. \& Chen, L. H. Thousandfold change in resistivity in
    magnetoresistive La-Ca-Mn-O films. \emph{Science} \textbf{264}, 413-415 (1994).
\bibitem{Shimakawa1996} Shimakawa, Y., Kubo, Y. \& Manako, T. Giant magnetoresistance in Tl$_2$Mn$_2$O$_7$ with the pyrochlore structure. \emph{Nature} \textbf{379}, 53–55 (1996).
\bibitem{Majumdar1998} Majumdar, P. \& Littlewood, P. B. Dependence of magnetoresistivity on charge-carrier density in metallic ferromagnets and doped magnetic semiconductors. \emph{Nature} \textbf{395}, 479–481 (1998).
\bibitem{Xiong2004} Xiong, Z. H., Wu, D., Valy Vardeny, Z. \& Shi, J. Giant magnetoresistance in organic spin-valves. \emph{Nature} \textbf{427}, 821–824 (2004).
\bibitem{Urushibara1995} Urushibara, A., Moritomo, Y., Arima, T., Asamitsu, A., Kido, G., \& Tokura, Y. Insulator-metal transition and giant magnetoresistance in La$_{1-x}$Sr$_x$MnO$_3$. \emph{Phys. Rev. B} \textbf{51}, 14103-14109 (1995).
\bibitem{Gladyshevskii1964} Gladyshevskii, E. I. \& Kripyakevich, P. I. Monosilicides of rare-earth metals and their crystal structures. \emph{Zh. Strukt.
    Khim.} \textbf{5}, 853-859 (1964).
\bibitem{nagaki1990} Nagaki, D. A. \& Simon, A. Structure of gadolinium monosilicide. \emph{Acta Cryst.} \textbf{C46}, 1197-1199 (1990).
\bibitem{Saito1996} Saito, H., Suzuki, S., Fukamichi, K., Mitamura, H. \& Goto, T. Metamagnetic transition in GdSi. \emph{J. Phys. Soc. Jpn.} \textbf{65},
    1938-1940 (1996).
\bibitem{Tung2005} Tung, L. D., Lees, M. R., Balakrishnan, G., Paul, D. McK., Schobinger-Papamantellos, P., Tegus, O., Brommer, P. E., \& Buschow, K. H. J.
    Field-induced magnetic phase transitions in a GdSi single crystal. \emph{Phys. Rev. B} \textbf{71}, 144410-4 (2005).
\bibitem{Roger2006} Roger, J., Babizhetskyy, V., Hiebl, K., Halet, J.-F. \& Gu\'{e}rin, R. Structural chemistry, magnetism and electrical properties of
    binary Gd silicides and Ho$_3$Si$_4$. \emph{J. Alloys Compd.} \textbf{407}, 25-35 (2006).
\bibitem{Heikes1964} Heikes, R. R. \& Chen, C. W. Evidence for impurity bands in La-doped EuS. \emph{Physics}, \textbf{1}, 159-160 (1964).
\bibitem{Barrera2005} Barrera, G. D., Bruno, J. A. O., Barron, T. H. K. \& Allan, N. L. Negative thermal expansion. \emph{J. Phys. Condens. Matter}
    \textbf{17}, R217-R252 (2005).
\bibitem{Ibarra1995} Ibarra, M. R., Algarabel, P. A., Marquina, C., Blasco, J. \& Garc\'{i}a, J. Large magnetovolume effect in yttrium doped La-Ca-Mn-O
    perovskite. \emph{Phys. Rev. Lett.} \textbf{75}, 3541-3544 (1995).
\bibitem{Castilho1991} Castilho, J. H., Chambouleyron, I., Marques, F. C., Rettori, C. \& Alvarez, F. Electrical conductivity of amorphous silicon doped
    with rare-earth elements. \emph{Phys. Rev. B} \textbf{43}, 8946-8950 (1991).
\bibitem{Hellman1996} Hellman, F., Tran, M. Q., Gebala, A. E., Wilcox, E. M. \& Dynes, R. C. Metal-insulator transition and giant negative
    magnetoresistance in amorphous magnetic rare earth silicon alloys. \emph{Phys. Rev. Lett.} \textbf{77}, 4652-4655 (1996).
\bibitem{Hellman2000} Hellman, F., Queen, D. R., Potok, R. M. \& Zink, B. L. Spin-glass freezing and RKKY interactions near the metal-insulator transition
    in amorphous Gd-Si alloys. \emph{Phys. Rev. Lett.} \textbf{84}, 5411-5414 (2000).
\bibitem{Kusz2000} Kusz, J., B$\ddot{\texttt{o}}$hm, H. \& Talik, E. X-ray investigation and discussion of the magnetostriction of Gd$_3$\emph{T} (\emph{T} = Ni, Rh,
    Ir$_x$) single crystals. \emph{J. Appl. Cryst.} \textbf{33}, 213-217 (2000).
\bibitem{Mackintosh1962} Mackintosh, A. R. Magnetic ordering and the electronic structure of rare-earth metals. \emph{Phys. Rev. Lett.} \textbf{9},
    90-93 (1962).
\bibitem{Yamada1974} Yamada, H. \& Takada, S. On the electrical resistivity of antiferromagnetic metals at low temperatures. \emph{Prog. Theor. Phys.}
    \textbf{52}, 1077-1093 (1974).
\bibitem{Mackintosh1964} Mackintosh, A. R. \& Spanel, L. E. Magnetoresistance in rare earth single crystals. \emph{Solid State Commun.} \textbf{2},
    383-386 (1964).
\bibitem{Yamada1973} Yamada, H. \& Takada, S. Magnetoresistance of antiferromagnetic metals due to \emph{s-d} interactions. \emph{J. Phys. Soc. Jpn.}
    \textbf{34}, 51-57 (1973).
\bibitem{Jensen1991} Jensen, J. \& Mackintosh, A. R. \emph{Rare Earth Magnetism: Structures and Excitations} (Clarendon Press, Oxford, 1991).
\bibitem{Moln2007} Moln$\acute{a}$r, S. \& Stampe, P. A. \emph{Magnetic polarons in "Handbook of magnetism and advanced magnetic materials".} (John Wiley \& Sons, Ltd, 2007).
\bibitem{Yosida1957} Yosida, K. Anomalous electrical resistivity and magnetoresisitane due to an \emph{s}-\emph{d} interactin in Cu-Mn Alloys. \emph{Phys. Rev.} \textbf{107}, 396-403 (1957).
\bibitem{Bogach2006} Bogach, A. V., Burkhanov, G. S., Chistyakov, O. D., Glushkov, V. V., Demishev, S. V., Samarin, N. A., Paderno, Yu. B., Shitsevalova, N. Yu. \& Sluchanko, N. E. Bulk and local magnetization in CeAl$_6$ and CeB$_6$. Physica B \textbf{378-380}, 769-770 (2006).
\bibitem{Anisimov2009} Anisimov, M. A., Bogach, A. V., Glushkov, V. V., Demishev, S. V., Samarin, N. A., Shitsevalova, N. Yu. \& Sluchanko, N. E. Low temperature magnetotransport in RB$_6$ (R = Pr, Nd). \emph{J. Phys.: Conf. Ser.} \textbf{150}, 042005-4 (2009).
\bibitem{Mott1936} Mott, N. F. \& Wills, H. H. The electrical conductivity of transition metals. \emph{Proc. R. Soc. Lond. A} \textbf{153}, 699-717 (1936).
\bibitem{Wilson1938} Wilson, A. H. The electrical conductivity of the transition metals. \emph{Proc. R. Soc. Lond. A} \textbf{167}, 580-593 (1938).
\bibitem{Singh1991} Singh, D. J. Adequacy of the local-spin-density approximation for Gd. \emph{Phys. Rev. B} \textbf{44}, 7451-7454 (1991).
\bibitem{Rodriguez-Carvajal1993} Rodr\'{\i}guez-Carvajal, J. Recent advances in magnetic structure determination by neutron powder diffraction. Physica
    B \textbf{192}, 55-69 (1993).
\end{thebibliography}

\begin{thebibliography}{9}
\bibitem{Rodriguez-Carvajal1993} Rodr\'{\i}guez-Carvajal, J. Recent advances in magnetic structure determination by neutron powder diffraction. Physica
    B \textbf{192}, 55-69 (1993).
\bibitem{Haifeng2009} Li, H., Tian, W., Zarestky, J. L., Kreyssig, A., Ni, N., Bud$'$ko, S. L., Canfield, P. C., Goldman, A. I., McQueeney, R. J. \&
    Vaknin, D. Magnetic and lattice coupling in single-crystal SrFe$_2$As$_2$: A neutron scattering study. \emph{Phys. Rev. B} \textbf{80}, 054407-5 (2009).
\bibitem{Fawcett1988} Fawcett, E. Spin-density-wave antiferromagnetism in chromium. \emph{Rev. Mod. Phys.} \textbf{60}, 209-283 (1988).
\bibitem{Hemberger2006} Hemberger, J., Rudolf, T., Krug von Nidda, H.-A., Mayr, F., Pimenov, A., Tsurkan, V. \& Loidl, A. Spin-driven phonon splitting in
    bond-frustrated ZnCr$_2$S$_4$. \emph{Phys. Rev. Lett.} \textbf{97}, 087204-4 (2006).
\bibitem{Tung2005} Tung, L. D., Lees, M. R., Balakrishnan, G., Paul, D. McK., Schobinger-Papamantellos, P., Tegus, O., Brommer, P. E., \& Buschow, K. H. J.
    Field-induced magnetic phase transitions in a GdSi single crystal. \emph{Phys. Rev. B} \textbf{71}, 144410-4 (2005).
\bibitem{Yamada1974} Yamada, H. \& Takada, S. On the electrical resistivity of antiferromagnetic metals at low temperatures. \emph{Prog. Theor. Phys.}
    \textbf{52}, 1077-1093 (1974).
\bibitem{Roger2006} Roger, J., Babizhetskyy, V., Hiebl, K., Halet, J.-F. \& Gu\'{e}rin, R. Structural chemistry, magnetism and electrical properties of
    binary Gd silicides and Ho$_3$Si$_4$. \emph{J. Alloys Compd.} \textbf{407}, 25-35 (2006).
\end{thebibliography}
\end{document}